\documentclass{ws-mpla}
\usepackage[super]{cite}
\usepackage{graphicx}
\begin{document}

\markboth{G.~L.~Klimchitskaya}
{The Casimir-Polder Interaction of an Atom and Real Graphene
Sheet}

\catchline{}{}{}{}{}

\title{\uppercase{The Casimir-Polder interaction of an atom and real graphene
sheet: Verification of the Nernst heat theorem}
}

\author{\uppercase{ G.~L.~Klimchitskaya}}

\address{Central Astronomical Observatory at Pulkovo of the
Russian Academy of Sciences, Saint Petersburg,
196140, Russia\\
Institute of Physics, Nanotechnology and
Telecommunications, Peter the Great Saint Petersburg
Polytechnic University, Saint Petersburg, 195251, Russia\\
g\_klimchitskaya@mail.ru}

\maketitle

\pub{Received 12 July 2019}{Revised 12 August 2019}

\begin{abstract}
We find the low-temperature behavior of the Casimir-Polder free energy
and entropy for an atom interacting with real graphene sheet
possessing nonzero energy gap and chemical potential. Employing the
formalism of the polarization tensor, it is shown that the
Casimir-Polder entropy goes to zero by the power law with vanishing
temperature, i.e., the Nernst heat theorem is satisfied. This result
is discussed in connection with the problems connected with account of
free charge carriers in the Lifshitz theory.

\end{abstract}

\ccode{PACS Nos.: 12.20.Ds, 34.35.+a, 72.80.Vp}

\section{Introduction}	

During the last two decades the Nernst heat theorem was repeatedly used
as a test for models of dielectric response in theory of Casimir and
Casimir-Polder interactions.\cite{1} Specifically it was shown that for
metallic test bodies with perfect crystal lattices (two parallel plates,
thin films or a sphere above a plate) the low-temperature behavior
of the Casimir free energy and entropy satisfies this theorem if the
relaxation properties of free electrons are disregarded and violates
it if they are accounted for.\cite{2,3,4} For dielectric plates the
Nernst heat theorem is satisfied if the dc conductivity of plate
materials is omitted and violated otherwise\cite{5,6} (see
Ref.~\citen{7} for a review). These results were generalized for the
case of magnetic metals\cite{8} and ferromagnetic dielectrics\cite{9}.
It was also shown that the Casimir-Polder interaction of an atom with
a dielectric plate satisfies and violates the Nernst theorem if the
dc conductivity of plate material is omitted and included,
respectively.\cite{10} The attempt\cite{11} to avoid inconsistencies
with thermodynamics by including the screening effects and diffusion
currents in the Lifshitz theory turned out to be unsuccessful.\cite{12,13}

In all above cases, the dielectric response of the test body material
is described by some phenomenological expression, e.g., by the
plasma or Drude model. Because of this, it is of much interest to
consider thermodynamic properties of the Casimir and Casimir-Polder
interactions for graphene sheets whose dielectric response was
found on the basis of first principles of quantum electrodynamics at
nonzero temperature in the framework of Dirac model. Thus, it was
shown that for two parallel pristine (perfect) graphene sheets the
Casimir entropy satisfies the Nernst theorem.\cite{14} The same is true
for the Casimir-Polder entropy of an atom interacting with a pristine
graphene sheet.\cite{15} In both cases the leading terms in the
low-temperature expansions of the free energy are
determined by an explicit parametric temperature dependence of the
polarization tensor  which is the fundamental quantity
describing the response of graphene to electromagnetic field.

In this paper, we find the low-temperature behavior of the Casimir-Polder
free energy and entropy for an atom interacting with real graphene sheet
possessing a nonzero energy gap $\Delta$ and chemical potential $\mu$
satisfying the condition $\Delta > 2\mu$. It is shown that for a
real graphene sheet the leading term in the low-temperature expansion
of the free energy is not the same as for a pristine graphene. It is
determined by a summation over the Matsubara frequencies of the
zero-temperature part of the polarization tensor rather than by a
dependence of this tensor on temperature as a parameter. Although
our results are again in agreement with the Nernst heat theorem, they
do not support the statement\cite{16} that for the case
$\Delta > 2\mu$ the leading term in the Casimir-Polder free energy
vanishes with $T$ exponentially fast.

The paper is organized as follows. In Sec. 2 we present the general
formalism in terms of the polarization tensor. Section 3 is devoted
to the low-temperature expansion of the Casimir-Polder free energy for
gapped graphene with $\Delta > 2\mu$. Section 4 contains
conclusions and discussion.

\section{The Casimir-Polder Interaction in Terms of the Polarization
Tensor at Low Temperature}
\def\p0l{\widetilde{\Pi}_{00,l}}
\def\pl{\widetilde{\Pi}_{l}}
\def\v{\tilde{v}_{\rm F}}
\def\rme{\mathrm{e}}

The Casimir-Polder free energy of
an atom interacting with real graphene sheet
spaced at a distance $a$ in thermal equilibrium with the environment at
temperature $T$ can be written in terms of dimensionless variables as follows:\cite{1}
\begin{equation}
{\cal F}(a,T)=-\frac{k_BT}{8a^3}\sum_{l=0}^{\infty}{\vphantom{\sum}}^{\prime}\!
\alpha_l\!\int_{\zeta_l}^{\infty}\!\!\!dy\,\rme^{-y}\left[(2y^2-\zeta_l^2)r_{\rm TM}(i\zeta_l,y)
-\zeta_l^2r_{\rm TE}(i\zeta_l,y)\right].
\label{eq1}
\end{equation}
\noindent
Here, $k_B$ is the Boltzmann constant, $\zeta_l=\xi_l/\omega_c$,
$\xi_l=2\pi k_BTl/\hbar$ ($l=0,\,1,\,2,\,\ldots$) are the Matsubara frequencies,
$\omega_c=c/(2a)$ is the characteristic frequency, and
$\alpha_l\equiv\alpha(i\omega_c\zeta_l)$ are the values of atomic electric
polarizability at the frequencies $i\xi_l$ (the prime on the summation sign
divides the term with $l=0$ by 2).

The reflection coefficients for two polarizations of the electromagnetic field,
transverse magnetic (TM) and transverse electric (TE), on graphene are expressed
via the dimensionless polarization tensor $\widetilde\Pi_{mn}$ as\cite{17}
\begin{equation}
r_{\rm TM}(i\zeta_l,y)=\frac{y\p0l}{y\p0l+2(y^2-\zeta_l^2)},
\quad
r_{\rm TE}(i\zeta_l,y)=-\frac{\pl}{\pl+2y(y^2-\zeta_l^2)}.
\label{eq2}
\end{equation}
\noindent
Here, $\widetilde\Pi_{mn,l}=\widetilde\Pi_{mn}(i\zeta_l,y)=
2a\Pi_{nm}(i\xi_l,k_{\bot})/\hbar$, where $\Pi_{mn}\ (m,\,n=0,\,1,\,2)$ is the
dimensional polarization tensor of graphene, $k_{\bot}$ is the magnitude
of the wave vector projection on its plane so that
$y=2a(k_\bot^2+\xi_l^2/c^2)^{1/2}$, $\widetilde\Pi_{{\rm tr},l}=\widetilde\Pi_{m,l}^{\,\,m}$,
and
\begin{equation}
\pl=(y^2-\zeta_l^2)\widetilde\Pi_{{\rm tr},l}-y^2\p0l.
\label{eq3}
\end{equation}

The polarization tensor of gapped graphene was first found\cite{18,19} at the
frequencies $i\xi_l$ and then generalized for the entire complex plane.\cite{20}
The results of Ref.~\citen{20} were extended for the case of graphene with
nonzero chemical potential.\cite{21}
This tensor was used to describe the
Casimir effect,\cite{14,15,16,17,22,23,24,25,26,31,32}
reflectances\cite{33,34,35,36,37} and conductivity
properties\cite{38,39,40,41} in graphene systems.
Eventually the formalisms using the polarization tensor and the density-density
correlation functions in the random phase approximation are equivalent.\cite{25}
In Ref.~\citen{17} the quantities $\p0l$ and $\pl$ were separated into the
parts independent and dependent on $\mu$
\begin{eqnarray}
&&
\p0l(y,T,\Delta,\mu)=\p0l^{(0)}(y,\Delta)+\p0l^{(1)}(y,T,\Delta,\mu),
\nonumber \\
&&
\pl(y,T,\Delta,\mu)=\pl^{(0)}(y,\Delta)+\pl^{(1)}(y,T,\Delta,\mu).
\label{eq4}
\end{eqnarray}
\noindent
The explicit expressions for the first contributions are given by\cite{17,18}
\begin{equation}
\p0l^{(0)}=\alpha\frac{y^2-\zeta_l^2}{p_l}\Psi(D_l), \quad
\pl^{(0)}=\alpha(y^2-\zeta_l^2){p_l}\Psi(D_l),
\label{eq5}
\end{equation}
\noindent
where $\alpha=e^2/(\hbar c)$ is the fine-structure constant,
$D_l=2a\Delta/(\hbar cp_l)$,
\begin{equation}
\Psi(x)=2[x+(1-x^2)\arctan(x^{-1})],\quad
p_l=\sqrt{\v^2y^2+(1-\v^2)\zeta_l^2}
\label{eq6}
\end{equation}
\noindent
and $\v=v_{\rm F}/c\approx 1/300$ is the dimensionless Fermi velocity.

The second contributions on the right-hand side of Eq.~(\ref{eq4})
have the form
\begin{eqnarray}
&&
\p0l^{(1)}=\frac{4\alpha p_l}{\v^2}\int_{D_l}^{\infty}\!\!\!
du\,w_l(u,y,T,\mu)X_{00,l}(u,y,T),
\nonumber \\
&&
\pl^{(1)}=-\frac{4\alpha p_l}{\v^2}\int_{D_l}^{\infty}\!\!\!
du\,w_l(u,y,T,\mu)X_{l}(u,y,T),
\label{eq7}
\end{eqnarray}
\noindent
where
\begin{equation}
w_l=\left[\rme^{B_lu+\frac{\mu}{k_BT}}+1\right]^{-1}+
\left[\rme^{B_lu-\frac{\mu}{k_BT}}+1\right]^{-1},
\label{eq8}
\end{equation}
\noindent
$B_l=\hbar cp_l/(4ak_BT)$ and the quantities $X_{00,l}$ and $X_{l}$ are
defined as\cite{17}
\begin{eqnarray}
&&
X_{00,l}(u,y,T)=1-{\rm Re}\frac{p_l-p_lu^2+2i\zeta_lu}{[p_l^2-p_l^2u^2+
\v^2(y^2-\zeta_l^2)D_l^2+2i\zeta_lp_lu]^{1/2}},
\label{eq9}\\
&&
X_{l}(u,y,T)=\zeta_l^2-p_l{\rm Re}\frac{\zeta_l^2-p_l^2u^2+\v^2(y^2-\zeta_l^2)D_l^2
+2i\zeta_lp_lu}{[p_l^2-p_l^2u^2+
\v^2(y^2-\zeta_l^2)D_l^2+2i\zeta_lp_lu]^{1/2}}.
\nonumber
\end{eqnarray}

Note that in the limiting case $\mu\to 0$ the quantities $\p0l^{(1)}$ and  $\pl^{(1)}$
have the meaning of thermal corrections to the  polarization tensor at $T=0$
and, thus, go to zero when $T$ vanishes. For $\mu\neq 0$ this may be and may be not so
depending on the value of $\Delta$. Specifically, it was proven\cite{17} that under
the condition $\Delta\geq 2\mu$  the quantities
$\p0l^{(1)}$ and  $\pl^{(1)}$  again vanish with vanishing $T$ and, thus, play the role
of thermal corrections. Then $\p0l(T=0)=\p0l^{(0)}$ which does not depend on $\mu$
and $\p0l^{(1)}=\delta_T\p0l$. In a similar way, $\pl(T=0)=\pl^{(0)}$
and $\pl^{(1)}=\delta_T\pl$ in this case.

Now we take into account that $\delta_T\p0l/\p0l(T=0)$ and $\delta_T\pl/\pl(T=0)$
go to zero with vanishing $T$. Expanding the reflection coefficients (\ref{eq2}) in powers
of these parameters and preserving only the first-order terms, one obtains
\begin{equation}
r_{\rm TM(TE)}(i\zeta_l,y)=r_{\rm TM(TE)}^{(0)}(i\zeta_l,y)+
\delta_Tr_{\rm TM(TE)}(i\zeta_l,y),
\label{eq13}
\end{equation}
\noindent
where
\begin{equation}
r_{\rm TM}^{(0)}(i\zeta_l,y)=\frac{y\p0l(T=0)}{y\p0l(T=0)+2(y^2-\zeta_l^2)},
\quad
r_{\rm TE}^{(0)}(i\zeta_l,y)=-\frac{\pl(T=0)}{\pl(T=0)+2y(y^2-\zeta_l^2)}
\label{eq14}
\end{equation}
\noindent
and
\begin{eqnarray}
&&
\delta_Tr_{\rm TM}(i\zeta_l,y)=\frac{2y(y^2-\zeta_l^2)\delta_T\p0l}{[y
\p0l(T=0)+2(y^2-\zeta_l^2)]^2},
\label{eq15}\\
&&
\delta_Tr_{\rm TE}(i\zeta_l,y)=-\frac{2y(y^2-\zeta_l^2)\delta_T\pl}{[\pl(T=0)+
2y(y^2-\zeta_l^2)]^2}.
\nonumber
\end{eqnarray}

Using Eqs.~(\ref{eq1}) and (\ref{eq13})--(\ref{eq15}) one can find the asymptotic
behavior of the Casimir-Polder free energy at arbitrarily low temperature.

\section{Low-Temperature Behavior of the Casimir-Polder Free Energy and Entropy}

We consider first graphene with $\Delta\geq 2\mu$ and present the temperature-dependent
part of the free energy (\ref{eq1}) in the form
\begin{equation}
\delta_T{\cal F}(a,T)=\delta_T^{(1)}{\cal F}(a,T)+\delta_T^{(2)}{\cal F}(a,T),
\label{eq16}
\end{equation}
\noindent
where $\delta_T^{(1)}{\cal F}(a,T)$ is obtained with reflection coefficients (\ref{eq14})
and $\delta_T^{(2)}{\cal F}(a,T)$ with the thermal corrections to them (\ref{eq15}).
It is evident that the temperature dependence  of $\delta_T^{(1)}{\cal F}(a,T)$
originates exclusively from the summation over the Matsubara frequencies in the Lifshitz
formula, whereas $\delta_T^{(2)}{\cal F}(a,T)$ also depends on $T$ via an explicit dependence
of the polarization tensor on $T$ as a parameter.

In was shown\cite{15} that using the Abel-Plana formula one can write
\begin{equation}
\delta_T^{(1)}{\cal F}(a,T)=-i\frac{\alpha_0k_BT}{8a^3}\int_0^{\infty}
dt\frac{\Phi(it\tau)-\Phi(-it\tau)}{\rme^{2\pi t}-1},
\label{eq17}
\end{equation}
\noindent
where $\tau=4\pi ak_BT/(\hbar c)$ and
$\Phi=\Phi_1+\Phi_2$ with $\Phi_1,\,\,\Phi_2$ defined as
\begin{eqnarray}
&&
\Phi_1(x)=2\int_x^{\infty}\!\!dy\,y^2\rme^{-y}r_{\rm TM}^{(0)}(ix,y),
\label{eq18} \\
&&
\Phi_2(x)=-x^2\int_x^{\infty}\!\!dy\,y^2\rme^{-y}\left[r_{\rm TM}^{(0)}(ix,y)+
r_{\rm TE}^{(0)}(ix,y)\right].
\nonumber
\end{eqnarray}
\noindent
Note that we are looking for the leading term in the expansion of ${\cal F}$ at low $T$.
Because of this, in Eq.~(\ref{eq17}) only the first term $\alpha_0$ in the expansion of
$\alpha(it\tau)=\alpha(ix)$ in powers of $x$ is preserved.

An expansion of $\Phi_1$ in the Taylor series in powers of $x$ is made using Eqs.~(\ref{eq5})
and (\ref{eq14}) where $\zeta_l$ should be replaced with $\tau t=x$. In so doing all
integrals are convergent. Direct calculation shows that
$\Phi_1^{\prime}(0)=\Phi_1^{(3)}(0)=\Phi_1^{(5)}(0)=0$.
Taking into account that the even powers in $x$ do not contribute to Eq.~(\ref{eq17}),
one concludes that the leading contribution of $\Phi_1$ to $\delta_T^{(1)}{\cal F}$
is of higher order in $T$ than $T^6$.

The function $\Phi_2$ cannot be expanded in the Taylor series around the point $x=0$.
To find its behavior at small $x$, we use the following expansion
\begin{equation}
r_{\rm TM}^{(0)}(ix,y)+r_{\rm TE}^{(0)}(ix,y)=r_{\rm TM}^{(0)}(0,y)+
r_{\rm TE}^{(0)}(0,y)+\frac{x^2}{2}\left[r_{\rm TM}^{(0)}(ix,y)+
r_{\rm TE}^{(0)}(ix,y)\right]_{x=0}^{\prime\prime}.
\label{eq19}
\end{equation}
\noindent
Here, the terms of order $x$ vanish because $r_{\rm TM(TE)}^{(0)}$ are the functions of $x^2$.
Direct calculation using Eqs.~(\ref{eq5})
and (\ref{eq14}) shows that the contribution of
$r_{\rm TM}^{(0)}(0,y)+r_{\rm TE}^{(0)}(0,y)$ to $\Phi_2$ takes the form
$b_1x^2+b_2x^4+O(x^5)$. The respective contribution to $\delta_T^{(1)}{\cal F}$
is of the order of $T^6$ or higher.

As a result, the leading contribution to $\Phi_2$ is
\begin{eqnarray}
&&
\Phi_2(x)=-\frac{x^4}{2}\int_x^{\infty}\!\!dy\,\rme^{-y}\left[r_{\rm TM}^{(0)}(ix,y)+
r_{\rm TE}^{(0)}(ix,y)\right]_{x=0}^{\prime\prime}
\nonumber \\
&&~~~
=\frac{\hbar c\alpha(1+\v^2)}{3\v^2a\Delta}x^4{\rm Ei}(-x)+Cx^4+O(x^5),
\label{eq20}
\end{eqnarray}
\noindent
where ${\rm Ei}(z)$ is the exponential integral function and
$C$ is the constant. As a result
\begin{equation}
\Phi(i\tau t)-\Phi(-i\tau t)=-i\frac{\pi\hbar c\alpha(1+\v^2)}{3\v^2a\Delta}
(\tau t)^4
\label{eq21}
\end{equation}
\noindent
and the leading contribution to the free energy due to  summation over
$\zeta_l$ is
\begin{equation}
\delta_T^{(1)}{\cal F}(a,T)=-\frac{\alpha_0(k_BT)^5}{(\hbar c)^3\Delta}
\frac{8\alpha(1+\v^2)}{\v^2}.
\label{eq22}
\end{equation}

Now we consider the second contribution to the thermal correction
$\delta_T^{(2)}{\cal F}$ which is expressed via the reflection coefficients (\ref{eq15})
and, finally, via $\p0l^{(1)}$ and $\pl^{(1)}$ defined in Eqs.~(\ref{eq7})--(\ref{eq9}).
Taking into account\cite{17} that the functions $X_{00,l}$ and $X_l$ behave as powers
of $T$ at small $T$, the behavior of $\p0l^{(1)}$ and $\pl^{(1)}$ is determined by
the exponential functions in (\ref{eq8}). In doing so, the major contribution is given
by ${\rm exp}[B_lu-\mu/(k_BT)]$. Note that at $u=D_l$
one has $B_lu=\Delta/(2k_BT)$. Because of this, the minimum value of the exponent
power $B_lu-\mu/(k_BT)$ is equal to $(\Delta-2\mu)/(2k_BT)$.

Basing on this, we deal below with the strict inequality $\Delta>2\mu$ and
temperature satisfying the condition $k_BT\ll\Delta-2\mu$. In this case
${\rm exp}[(\Delta-2\mu)/(2k_BT)]\gg 1$ and one can neglect by unity in the
denominator of the main second term in Eq.~(\ref{eq8}).
Introducing the new variable
$t=u-D_l$, we rewrite  Eq.~(\ref{eq7}) in the form
\begin{eqnarray}
&&
\p0l^{(1)}\approx\rme^{-\frac{\Delta-2\mu}{2k_BT}}
\frac{4\alpha p_l}{\v^2}\int_{0}^{\infty}\!\!\!
dt\,\rme^{-B_lt}X_{00,l}(t+D_l,y,T),
\nonumber \\
&&
\pl^{(1)}\approx-\rme^{-\frac{\Delta-2\mu}{2k_BT}}
\frac{4\alpha p_l}{\v^2}\int_{0}^{\infty}\!\!\!
dt\,\rme^{-B_lt}X_{l}(t+D_l,y,T).
\label{eq23}
\end{eqnarray}
\noindent
Remembering that in this case $\p0l^{(1)}=\delta_T\p0l$ and $\pl^{(1)}=\delta_T\pl$,
we conclude from Eqs.~(\ref{eq15}) and (\ref{eq23}) that
\begin{equation}
\delta_Tr_{\rm TM}(i\zeta_l,y)\sim \delta_Tr_{\rm TM}(i\zeta_l,y)\sim
\rme^{-\frac{\Delta-2\mu}{2k_BT}}.
\label{eq24}
\end{equation}

Taking into consideration that integration with respect to $t,\,\,y$
 and summation in $l$ cannot affect the result (\ref{eq24}) but may only add
 some power-type factors  of $k_BT$,
 one concludes that in the case $\Delta>2\mu$ , $k_BT\ll\Delta-2\mu$ it holds
 \begin{equation}
 \delta_T^{(2)}{\cal F}\sim \rme^{-\frac{\Delta-2\mu}{2k_BT}}.
\label{eq25}
\end{equation}

Thus, for $\Delta>2\mu$ the low-temperature dependence of the Casimir-Polder
free energy  is given by Eq.~(\ref{eq22}). Because of this, for the
respective entropy one has
\begin{equation}
S(T)=-\frac{\partial{\cal F}}{\partial T}=
\frac{40\alpha_0k_B(k_BT)^4}{(\hbar c)^3\Delta}
\frac{\alpha(1+\v^2)}{\v^2},
\label{eq26}
\end{equation}
which goes to zero with vanishing $T$, i.e., the Nernst heat theorem is satisfied.
Taking into account that Eq.~(\ref{eq22}) remains valid for graphene with $\Delta=2\mu$,
we find that even in this case the entropy cannot go the zero with
$T$ faster than (\ref{eq26}).

The obtained result disagrees with Ref.~\citen{16}. For a pristine
graphene this work finds
$\delta_T{\cal F}(a,T)=C(a)T^3$ at low $T$
with no explicit expression for the coefficient $C(a)$.
The same expression  with an explicitly determined coefficient $C$ was,
however, obtained earlier.\cite{15}
As to the case of gapped and doped graphene satisfying the condition $\Delta>2\mu$,
it was concluded\cite{16} that
$ \delta_T{\cal F}\sim \rme^{-\frac{\Delta-2\mu}{2k_BT}}$ at low $T$.
This behavior is, however, valid only for the part  of $\delta_T{\cal F}$
determined by an explicit dependence of the polarization
tensor on temperature as a parameter. Thus, in Ref.~\citen{16}, the temperature
dependence arising from a summation over the Matsubara frequencies using the
part of the polarization tensor at $T=0$ was not properly addressed.

\section{Conclusions and Discussion}

As noted in Sec.~1, the Nernst heat theorem is an important test for
the models of dielectric response used in Casimir physics. The results
of many experiments on measuring the Casimir and Casimir-Polder
interaction performed during the last 20 years were found in
contradiction with theoretical predictions of the Lifshitz theory
obtained using standard dielectric response functions such as the
Drude model (see Ref.~\citen{1} and more recent
results\cite{43,45,47,48,49}). The same dielectric functions
lead to violation of the Nernst theorem when they are used in
calculations of the Casimir and Casimir-Polder entropy by means of
the Lifshitz theory. Taking into account that the Drude model is well
confirmed in numerous experiments with real electromagnetic fields
possessing a nonzero strength, this may signify some deep difference
between real and fluctuating fields which is still not clearly
understood.

In the foregoing we have found the behavior of
the Casimir-Polder free energy and tntropy at low $T$ for a real graphene sheet
with nonzero $\Delta$ and $\mu$.
The dielectric response of graphene was determined on the basis of
first principles of quantum electrodynamics at $T\neq 0$
using the formalism of the polarization tensor.
The obtained Casimir-Polder entropy is in perfect agreement
with the Nernst heat theorem by decreasing to zero
as power of $T$. This result can be related to the
fact that the gradient of the Casimir force between an Au sphere and
graphene-coated substrate calculated using the
thermodynamically-consistent formalism of the polarization
tensor\cite{26} is also in good agreement with the measurement
data.\cite{50}

It is important to note that the same dielectric response function
expressed via the polarization tensor but along the real frequency
axis was used for theoretical description of
reflectances\cite{33,34,35,36,37} and electrical
conductivity\cite{38,39,40,41} of graphene in real electromagnetic
fields. This means that, unlike phenomenological dielectric functions,
the fundamental response functions may be equally applicable to
describe the reaction of a system to real and fluctuating
fields. This observation might be helpful for resolution of
the discussed above problems in Casimir physics.

\end{document}